\documentclass[11pt]{iopart}

\usepackage{graphicx}
\usepackage{color}
\usepackage{cite}

\begin{document}

\title{Hybridization-related correction to the jellium model for fullerenes}

\author{A V Verkhovtsev$^{1,2}$, R G Polozkov$^2$, V K Ivanov$^2$, A V Korol$^1$, 
A V Solov'yov$^1$\footnote{On leave from A.F. Ioffe Physical-Technical Institute, St. Petersburg, Russia}}

\address{$^1$ Frankfurt Institute for Advanced Studies, Ruth-Moufang-Str. 1, 60438 Frankfurt am Main, Germany}
\address{$^2$ St. Petersburg State Polytechnic University, Politekhnicheskaya ul. 29, 195251 St. Petersburg, Russia}

\ead{verkhovtsev@fias.uni-frankfurt.de}


\begin{abstract}
We introduce a new type of correction for a more accurate description of fullerenes within the 
spherically symmetric jellium model.
This correction represents a pseudopotential which originates from the comparison between an 
accurate {\it ab initio} calculation and the jellium model calculation.
It is shown that such a correction to the jellium model allows one to account, at least partly, 
for the sp$^2$-hybridization of carbon atomic orbitals. 
Therefore, it may be considered as a more physically meaningful correction as compared with a 
structureless square-well pseudopotential which has been widely used earlier.
\end{abstract}

\pacs{31.15.A-, 31.15.xr, 36.40.-c}


\section{Introduction}

Since the discovery of fullerenes by Kroto \emph{et al.} \cite{Kroto_1985_Nature.318.162} in 1985, 
these molecules have been the objects of intensive experimental and theoretical investigations 
(see, e.g., \cite{Sattler_Handbook_of_Nanophysics_2}).
At present, the investigation of fullerenes is active since they are proposed to be used in various 
fields of science and technology.
For instance, excitation of fullerenes, placed in a biological medium, by an external radiation or 
incident heavy ions may lead to an active generation of secondary electrons or reactive oxygen species. 
This allows fullerenes to be potentially used as sensitizers in photodynamic therapy
\cite{Fullerenes_photosensitizers_2008}.
A very important fundamental problem closely related to the aforementioned application is an adequate 
description of dynamic response of fullerenes to external fields or to the interaction with projectiles.
Processes of scattering of electrons, photons and heavy charged particles on various atomic clusters 
and fullerenes, in particular, have been actively studied during the past several decades 
(see, e.g., the review \cite{Solovyov_review_2005_IntJModPhys.19.4143} and references therein).
Not the least of the factors for a proper description of the dynamic response of a many-electron system 
is an adequate description of the ground- and excited-state (including excitation into continuum) 
properties of the system under study.

Contemporary software for quantum-chemical calculations (e.g., Gaussian 09 \cite{g09}) provides an 
accurate quantitative description of the ground state of many-particle systems (fullerenes, in particular),
and allows one to obtain information on geometrical and chemical properties of the system.
However, the description of dynamic properties, which play an important role in the process of 
photoionization, by means of such programs faces significant difficulties.
Dynamic properties (e.g., dynamic polarizability) are closely related to the response of a
many-electron system to an external electromagnetic field. 
In many cases, the properties are governed by a collective excitation of electrons and the formation 
of plasmon resonances in the excitation spectra \cite{Kaplan_1997_ZPhysD.40.375}.
In various systems plasmon resonances lie either below the ionization threshold (in metal clusters)
or above it (e.g., in fullerenes).
Out of these two classes of atomic clusters, only the optical response of metal clusters has been 
calculated so far with the help of quantum-chemical programs 
(see, e.g. \cite{Rubio_1996_PhysRevLett.77.247, Solovyov_2004_JPhysB.37.L137}).
Collective electron excitations in fullerenes, which lie in the continuous spectrum, have not been
described so far by means of quantum-chemical programs.
However, this can be achieved within simplified model approximations.
A minimum requirement is that these approximations must provide an accurate quantitative description 
of the ground-state features of the systems under study, in order to be applied to the investigation 
of the dynamic response and to the calculation of the photoabsorption 
(or, in particular, photoionization) spectrum.

One of the well-known and widely used approaches is based on the jellium model 
\cite{Ekardt_1984_PhysRevB.29.1558}.
It was applied frequently to the description of ground-state properties of metal clusters \cite{Ekardt_1984_PhysRevB.29.1558,Kharchenko_1994_PhysRevA.50.1459,Polozkov_2009_PhysRevA.79.063203}
and fullerenes \cite{Yannouleas_Landman_1994_ChemPhysLett.217.175,Yabana_Bertsch_1993_PhysScr.48.633},
as well as to the investigation of photoexcitation processes arising in these systems
\cite{Yabana_Bertsch_1993_PhysScr.48.633, Guet_Johnson_1992_PhysRevB.45.11283, 
Jankala_2011_PhysRevLett.107.183401, PuskaNiemenen_1993_PhysRevA.47.1181, 
WendinWastberg_1993_PhysRevB.48.14764, Ruedel_2002_PhysRevLett.89.125503, 
Kidun_2004_JPhysB.37.L321, Polozkov_2005_JPhysB.38.4341, Madjet_2008_JPhysB.41.105101}.

In \cite{Yannouleas_Landman_1994_ChemPhysLett.217.175,Pavlyukh_2010_PhysRevA.81.042515}, it was stated 
that the ground-state properties of fullerenes cannot be described properly by the standard jellium 
model which produces, in particular, unreliable values for the total energy \cite{Yannouleas_Landman_1994_ChemPhysLett.217.175}.
To avoid this, adding of {\it structureless} pseudopotential corrections was suggested \cite{Yannouleas_Landman_1994_ChemPhysLett.217.175}.
As a rule, a phenomenological square-well (SW) pseudopotential has been commonly used in the 
calculations 
\cite{Yannouleas_Landman_1994_ChemPhysLett.217.175, PuskaNiemenen_1993_PhysRevA.47.1181, 
Ruedel_2002_PhysRevLett.89.125503, Madjet_2008_JPhysB.41.105101}.
It was claimed that accounting for such a pseudopotential increases the accuracy of the jellium-based 
description \cite{PuskaNiemenen_1993_PhysRevA.47.1181} and, for instance, allows one to reproduce the 
experimental value of the first ionization potential of C$_{60}$ \cite{Madjet_2008_JPhysB.41.105101}.
Nonetheless, the applicability of the jellium model for fullerenes and the choice of parameters of the 
used SW pseudopotential have not been clearly justified so far from a physical viewpoint.

In this paper, we use another methodology and introduce a {\it structured} pseudopotential which 
originates from the comparison of an accurate {\it ab initio} calculation with the jellium-based one.
Using this pseudopotential as a correction to the standard jellium model, 
one can introduce effects of the sp$^2$-hybridization of carbon atomic orbitals into the jellium 
model and relate parameters of the model with the features of the system obtained from the more 
precise calculation.
By means of the presented pseudopotential, a relatively simple jellium model acquires more physical 
sense and parameters of the model obtain a clear physical justification.
Hereby, we confirm the relevance of using the jellium model for the description of fullerenes.
Investigating two molecules, C$_{60}$ and C$_{20}$, we show that the results obtained have a common 
origin and that they could be also extended to other highly symmetric fullerenes.

The atomic system of units, $m_e = |e| = \hbar = 1$, is used throughout the paper.

\section{Methods of investigation}
\subsection{Jellium model}

In this paper, the fullerenes C$_{60}$ and C$_{20}$ are treated within the jellium model which is based 
on an assumption that a many-electron system is considered as a sum of two interacting subsystems: 
a valence electrons subsystem and a positively charged ionic core.
One of the stable isomers of C$_{20}$ corresponds geometrically to the regular dodecahedron
\cite{Prinzbach_2000_Nature.407.60} and, like the truncated icosahedron C$_{60}$, has the symmetry of 
the $I_h$ point group which is very close to spherical symmetry.
Therefore, a detailed ionic structure of the systems under study is substituted by the uniform 
spherically symmetric distribution of the positive charge, in the field of which the motion of the 
valence electrons is considered \cite{Ekardt_1984_PhysRevB.29.1558}.

The valence 2s$^2$2p$^2$ electrons in each carbon atom form a cloud of delocalized electrons, 
while the inner-shell 1s$^2$ electrons are treated as frozen and not taken into consideration.
Thereby, we consider 240 delocalized electrons in C$_{60}$ and 80 electrons in C$_{20}$.
The valence electrons are moving in a spherically symmetric central field, so one can construct
the electronic configuration described by the unique set of quantum numbers \{$ n,l$\} 
where $n$ and $l$ are the principal and orbital quantum numbers, respectively.

Since it is commonly acknowledged \cite{Zhang_Heath_Kroto_1986_JPhysChem.90.525,
Kratschmer_Fostiropoulos_1990_ChemPhysLett.170.167,Goroff_1996_JPhysChem.29.77} that C$_{60}$, 
as well as other fullerenes, is formed from fragments of planar graphite sheets, it is natural 
to match the $\sigma$- and $\pi$-orbitals of graphite to the nodeless and the single-node 
wavefunctions of a fullerene, respectively \cite{Martins_1991_ChemPhysLett.180.457}.
Carbon atoms within a graphite sheet are connected by $\sigma$-bonds, whereas different sheets 
are connected by $\pi$-bonds.
In the fullerene, the nodeless $\sigma$-orbitals are localized at the radius of the ionic core 
while the single-node $\pi$-orbitals are oriented perpendicularly to the fullerene surface.
The ratio of $\sigma$- to $\pi$-orbitals in C$_{60}$ should be equal to $3:1$ due to the
sp$^2$-hybridization of carbon orbitals \cite{Haddon_1986_ChemPhysLett.125.459}.
Thereby, the electronic configuration of the delocalized electrons in C$_{60}$ is written
in the form \cite{Yabana_Bertsch_1993_PhysScr.48.633}:
\begin{center}
1s$^2$2p$^6$3d$^{10}$4f$^{14}$5g$^{18}$6h$^{22}$7i$^{26}$8k$^{30}$9l$^{34}$10m$^{18}$ \\
2s$^2$3p$^6$4d$^{10}$5f$^{14}$6g$^{18}$7h$^{10}$.
\end{center}
\noindent 
Radial wavefunctions of the 1s $\dots$ 10m shells are nodeless, while the wavefunctions of the 
2s $\dots$ 7h shells have one radial node each.

Using the same methodology, one defines the electronic configuration of the 80 delocalized 
electrons in C$_{20}$ as follows:
\begin{center}
1s$^2$2p$^6$3d$^{10}$4f$^{14}$5g$^{18}$6h$^{10}$2s$^2$3p$^6$4d$^{10}$5f$^{2}$.
\end{center}

Within the jellium model the fullerene core of the charged carbon ions, C$^{4+}$, is described as 
a positively charged spherical layer of a finite thickness $\Delta R = R_2 - R_1$. 
The thickness $\Delta R$ is chosen to be equal to $1.5$ \AA \ which corresponds to a typical 
diameter of a carbon atom \cite{Ostling_1993_EurophysLett.21.539} and refers to experimental 
data from \cite{Ruedel_2002_PhysRevLett.89.125503}.
The potential of the core may be written as:
\begin{equation}
\fl
U_{\rm core}(r) = -N \times \left\{ \begin{array}{cl}
  \displaystyle{ \frac32 \frac{R_2^2 - R_1^2}
{R_2^3 - R_1^3}}, &  r < R_1 \\
 \displaystyle{ \frac{1}{2\left(R_2^3 - R_1^3 \right)}
\left(3R_2^2 - r^2\left(1 + \frac{2R_1^3}{r^3} \right)
\right) },
& R_1 \le r \le R_2\\
  \displaystyle{ \frac{1}{r} },
& r > R_2
             \end{array} \right.
  \ , \label{eq01}
\end{equation}

\noindent where $N$ is the number of delocalized electrons in a fullerene 
($N=240$ in C$_{60}$ and $N=80$ in C$_{20}$),
$R_1 = R -\Delta R/2$ and $R_2 = R + \Delta R/2$ with $R$ standing for a fullerene radius 
($R_{\rm C_{60}} = 3.54$ \AA \ and $R_{\rm C_{20}} = 2.04$ \AA \ 
\cite{Gianturco_2002_JChemPhys.116.2811}).

The electronic subsystem is treated within the local density approximation (LDA).
Single-electron wave functions $\phi_{nlm}({\bf r})$ and the corresponding energies $\varepsilon_{nl}$ 
are determined from a system of self-consistent Kohn-Sham equations:
\begin{equation}
\left[ -\frac{\Delta}{2} + U_{ \rm eff}(\textbf{r}) \right] \phi_{nlm}(\textbf{r}) = 
\varepsilon_{nl} \phi_{nlm}(\textbf{r}) \ ,
\end{equation}
\begin{equation}
U_{ \rm eff}(\textbf{r}) = 
U_{ \rm core}(r) + \int \frac{\rho(\textbf{r}')}{|\textbf{r}-\textbf{r}'|} d\textbf{r}' + 
U_{\rm XC}^{\rm LDA}(\textbf{r}) \ ,
\end{equation}
\begin{equation}
\rho(\textbf{r}) = \sum_{nl} \sum_{m=-l}^l \frac{N_{nl}}{2(2l+1)}|\phi_{nlm}(\textbf{r})|^2 \ ,
\end{equation}

\noindent where $N_{nl}$ is a number of electrons in the $nl$-shell.
The exchange-correlation potential $U_{\rm XC}^{\rm LDA}(\textbf{r})$ is represented as a sum of 
the Slater exchange potential and a correlation potential:
\begin{equation}
U_{\rm XC}^{\rm LDA}(\textbf{r}) = 
U_{\rm X}(\textbf{r}) + U_{\rm C}(\textbf{r}) = 
-\left( \frac{3}{\pi}\right)^{1/3}\rho^{1/3}(\textbf{r}) + U_{\rm C}(\textbf{r}) \ .
\end{equation}

\noindent 
In the calculations, we used Perdew and Zunger parameterization of the correlation potential \cite{Perdew_Zunger_1981_PhyRevB.23.5048} which is presented in the form
\begin{equation}
U_{\rm C}(r_s) = 
\varepsilon_{\rm C}(r_s) \frac{1 + 1.229 \sqrt{r_s} + 0.444 r_s}{1 + 1.053 \sqrt{r_s} + 0.333 r_s} \ ,
\end{equation}
\begin{equation}
\varepsilon_{\rm C}(r_s) = - \frac{0.142}{1 + 1.053 \sqrt{r_s} + 0.333 r_s} \ ,
\end{equation}

\noindent where $r_s(\textbf{r}) = (4\pi \rho({\bf r})/3)^{-1/3}$ is the local Wigner-Seitz radius
for the electronic subsystem and $\varepsilon_{\rm C}(r_s)$ is the correlation energy per electron.

\subsection{{\it Ab initio} calculations}

The {\it ab initio} calculations were performed using the Gaussian 09 package \cite{g09}. 
For the description of the C$_{60}$ and C$_{20}$ fullerenes we used the split-valence triple-zeta 
basis set 6-311+G(d) with an additional set of polarization and diffuse functions.
The systems were calculated by means of the density functional theory.
To account for the exchange and correlation corrections, the Slater exchange functional
\cite{Kohn_Sham_1965_PhyRev.140.A1133} and the local Perdew functional 
(the so-called Perdew Local, PL) \cite{Perdew_Zunger_1981_PhyRevB.23.5048} were used.
By applying these rather simple functionals we wanted to achieve a full similarity in the description
of the electronic subsystem within the jellium model and the {\it ab initio} approaches.

The total electrostatic potential of the system is represented as a sum of the nuclear and 
electronic parts:
\begin{equation}
U_{\rm tot}(\textbf{r}) = U_{\rm n}(\textbf{r}) + U_{\rm el}(\textbf{r}) = 
- \sum_A \frac{Z_A}{|\textbf{r} - \textbf{R}_A|} + 
\int \frac{\rho(\textbf{r}')}{|\textbf{r} - \textbf{r}'|}d\textbf{r}' \ .
\label{eq08}
\end{equation}

The electron density $\rho(\textbf{r})$ and the potential $U_{\rm n}(\textbf{r})$ created by all
carbon ions, C$^{4+}$(1s$^2$), were extracted from the Gaussian output file with the help of the
Multiwfn software \cite{Multiwfn}.
The potential $U_{\rm el}(\textbf{r})$ created by the delocalized electrons was calculated
separately using the extracted electron density.

The jellium model treats the fullerenes C$_{60}$ and C$_{20}$ as spherically symmetric objects
while a more precise {\it ab initio} calculation accounts for the real icosahedral symmetry
of the molecules.
Therefore, to draw an analogy between the two methods we averaged the exact electrostatic
potential and the electron density over the directions of the position vector ${\bf r}$:
\begin{eqnarray}
&\overline{U}_{\rm tot}(r)& = \overline{U}_{\rm n}(r) + \overline{U}_{\rm el}(r) \ , 
\nonumber \\
&\overline{U}_{\rm i}(r)& = \frac{1}{4\pi} \int U_{\rm i}({\bf r}) d\Omega \quad (\rm i = tot, n, el) \ , 
\nonumber \\
&\overline{\rho}(r)& = \frac{1}{4\pi} \int \rho({\bf r}) d\Omega \ .
\end{eqnarray}

\noindent 
The averaged electron density includes only delocalized electrons, while the inner electron
orbitals are excluded from the consideration.

\section{Numerical results}

In this section, we compare the results of the {\it ab initio} and the jellium model calculations
for C$_{60}$ and C$_{20}$. 
The fullerene C$_{60}$ is discussed in detail below.
The results for the C$_{20}$ molecule and the comparison with C$_{60}$ are discussed further in 
this section.


Using the methodology implemented in a number of papers 
\cite{Yannouleas_Landman_1994_ChemPhysLett.217.175, PuskaNiemenen_1993_PhysRevA.47.1181, 
Ruedel_2002_PhysRevLett.89.125503, Madjet_2008_JPhysB.41.105101},
we add a negative SW pseudopotential $U_{\rm SW}$ to the core potential (\ref{eq01}):
\begin{equation}
U_{\rm core}(r) \rightarrow \left\{ \begin{array}{cl}
 U_{\rm core}(r) + U_{\rm SW}
& \mbox{$,\ R_1 \le r \le R_2$} \\
  U_{\rm core}(r)
& \mbox{$,\ {\rm otherwise}$}
             \end{array} \ . \right.
\end{equation}

\noindent 
The depth of the SW potential was chosen to obtain the same value of the outer-shell ionization 
potential as the defined one from the quantum-chemical calculation.
The pseudopotential $U_{\rm SW}$ is shown by the dashed red curve in the lower panel of 
figure~\ref{figure3}.

\begin{figure}[h]
\centering
\includegraphics[scale=0.5,clip]{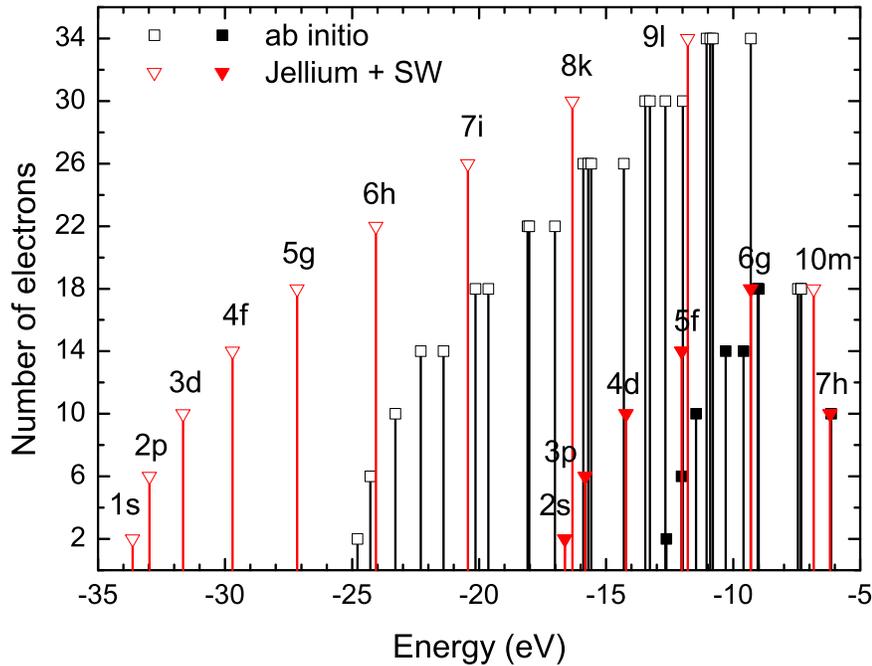}
\caption{
Single-electron energy levels of C$_{60}$ obtained from the {\it ab initio} calculation 
(empty and filled squares) and within the jellium model with an additional square-well (SW) 
pseudopotential (empty and filled triangles). 
Nodeless $\sigma$-orbitals and single-node $\pi$-orbitals are labeled by empty and filled symbols, 
respectively.}
\label{figure1}
\end{figure}

Single-electron energy spectra obtained from the {\it ab initio} calculation and within the jellium
model are presented in figure \ref{figure1}.
For the ease of perception, the height of levels of split stated (black lines) corresponds
to the occupation of shells within the jellium model (red lines).
Ionization potentials of several outer shells (6g, 10m and 7h) are in a good agreement with the
{\it ab initio} results while the remainder of the jellium spectrum is significantly broadened and 
differs from the more precise calculation.
It should be mentioned that none of the various jellium-based calculations of C$_{60}$ performed earlier
\cite{Yannouleas_Landman_1994_ChemPhysLett.217.175, PuskaNiemenen_1993_PhysRevA.47.1181, WendinWastberg_1993_PhysRevB.48.14764, Yabana_Bertsch_1993_PhysScr.48.633, Polozkov_2005_JPhysB.38.4341, Madjet_2008_JPhysB.41.105101} can produce the quantitative agreement of the single-electron
spectrum with that one obtained from the more precise {\it ab initio} calculation.

\begin{figure}[h]
\centering
\includegraphics[scale=0.5,clip]{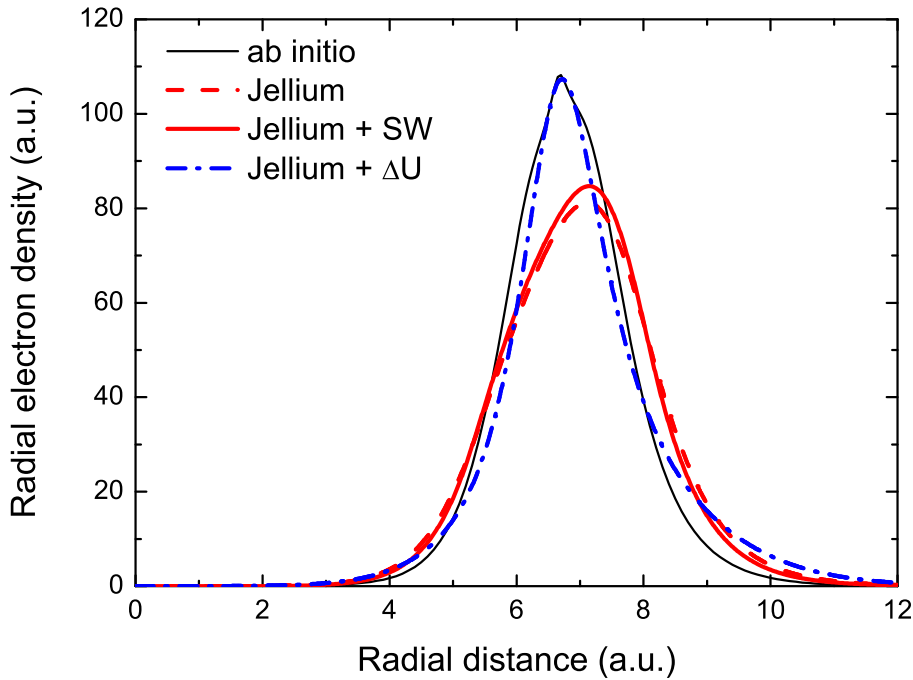}
\caption{
Radial electron density of C$_{60}$ obtained from the {\it ab initio} calculation (solid black curve) 
and calculated by means of the jellium model: the standard one (dashed red curve), with the additional 
SW pseudopotential (solid red curve) and with the additional pseudopotential $\Delta U$ 
(dash-dotted blue curve).}
\label{figure2}
\end{figure}

\begin{figure}[h]
\centering
\includegraphics[scale=0.35,clip]{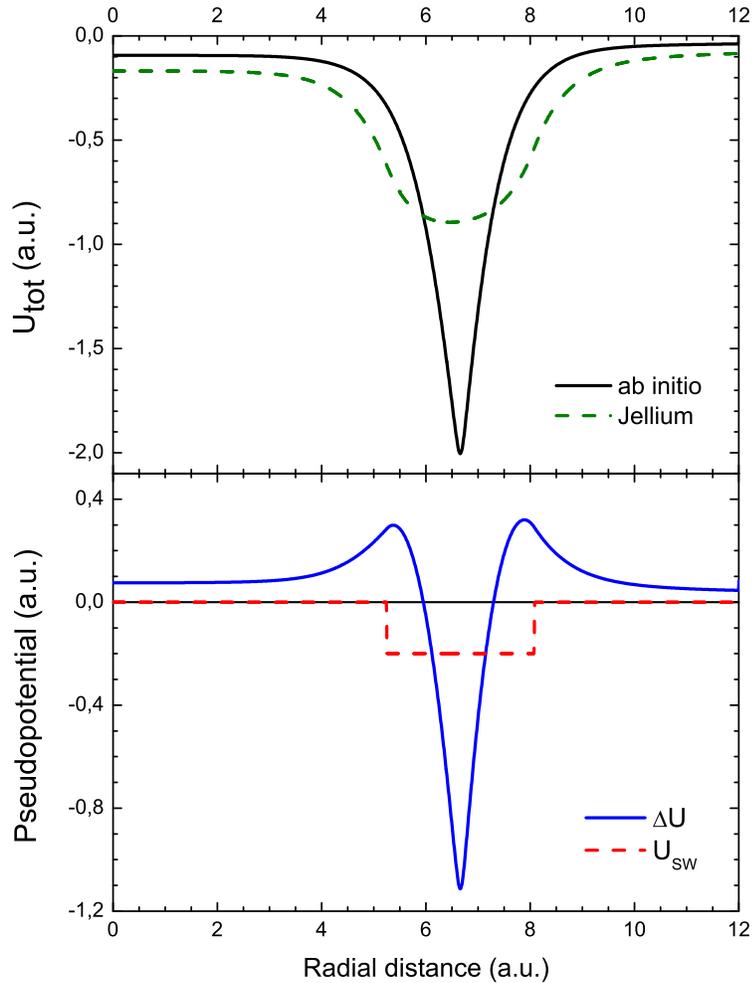}
\caption{
Upper panel: total electrostatic potential of C$_{60}$ obtained from the {\it ab initio} 
quantum chemistry calculation (solid curve) and within the jellium model (dashed curve). 
Lower panel: the difference $\Delta U$ between the total electrostatic potential of C$_{60}$ 
calculated by the {\it ab initio} methods and the one calculated within the jellium model 
(solid blue curve). 
The SW pseudopotential $U_{\rm SW}$ is also shown for the comparison (dashed red curve).}
\label{figure3}
\end{figure}

The radial density of the delocalized electrons obtained within the two approaches is presented 
in figure \ref{figure2}.
It is shown that the standard jellium model without any corrections (dashed red curve) fails to
represent the results of the {\it ab initio} calculation (black curve).
The additional SW pseudopotential does not modify the density distribution significantly
(solid red curve).

As shown in figures \ref{figure1} and \ref{figure2}, the jellium model with a simple additional
pseudopotential represents neither the single-electron energy spectrum nor the electron density 
distribution.
As opposed to more precise quantum chemistry methods, the jellium model does not take into account 
chemical features of the fullerene, such as hybridization of atomic orbitals in the formation of 
chemical bonding.
However, the jellium model can be improved by means of a more sophisticated pseudopotential which 
will allow one to describe chemical properties of the real system.
In this paper, we introduce the correction as a difference between the total electrostatic potential
of the system obtained from the {\it ab initio} calculation and the one obtained within the 
jellium model:
\begin{equation}
\Delta U(r) = U_{\rm tot}^{\rm QC}(r) - U_{\rm tot}^{\rm jel}(r) \ ,
\end{equation}

\noindent
where $U_{\rm tot}^{\rm QC}(r)$ is defined by Eq. (\ref{eq08}), and the potential 
$U_{\rm tot}^{\rm jel}(r)$ obtained within the jellium model is defined as:
\begin{equation}
U_{\rm tot}^{\rm jel}(r) = 
U_{ \rm core}(r) + \int \frac{\rho(\textbf{r}')}{|\textbf{r}-\textbf{r}'|} d\textbf{r}'  \ .
\end{equation}

\noindent 
The total potentials $U_{\rm tot}^{\rm QC}(r)$ and $U_{\rm tot}^{\rm jel}(r)$ of C$_{60}$ 
as well as their difference $\Delta U(r)$ are shown in figure \ref{figure3}.

We note that previously several spiritually close averaged pseudopotentials have been 
introduced to correct the jellium model.
For instance, it was done in the case of inhomogeneous electron gas on metal surfaces
\cite{Lang_Kohn_1970_PhysRevB.1.4555, Perdew_1976_PhysRevLett.37.1286}
and for spherically symmetric metallic clusters
\cite{Alasia_1995_PhysRevB.52.8488, Vieira_1997_JPhysB.30.3583}.

As opposed to the SW pseudopotential which affects equally all electrons of the system, 
$\Delta U$ is an alternating-sign pseudopotential (see the lower panel of figure \ref{figure3}), 
therefore it is attractive in the vicinity of the fullerene ionic core and repulsive at larger 
distances from the fullerene surface.
This means that such a potential affects differently the $\sigma$- and $\pi$-electrons of C$_{60}$ 
which are located on the surface of the molecule and perpendicularly to it, respectively.
Therefore, one can conclude that by means of a such potential it is possible to account,
to some extent, for the hybridization properties of the fullerene.

\begin{figure}[h]
\centering
\includegraphics[scale=0.5,clip]{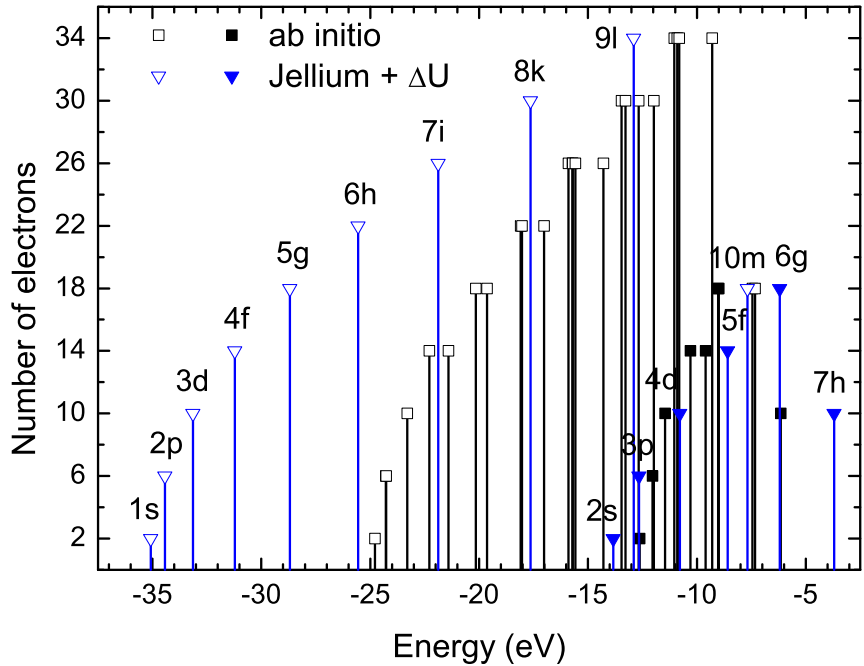}
\caption{
Single-electron energy levels of C$_{60}$ obtained from the {\it ab initio} calculation 
(empty and filled squares) and within the modified jellium model with an additional pseudopotential 
$\Delta U$ (empty and filled triangles). 
Nodeless $\sigma$-orbitals and single-node $\pi$-orbitals are labeled by empty and filled symbols, 
respectively.}
\label{figure5}
\end{figure}

The single-electron energy spectrum obtained within the ''modified'' jellium model with 
$\Delta U$ taken as an additional pseudopotential is presented in figure \ref{figure5}.
The modification allows one to obtain a better agreement of the jellium calculation with 
the {\it ab initio} one for the inner single-node 2s $\dots$ 5f orbitals.
On the contrary, it shifts the 6g and 7h ionization potentials by 2.8 and 2.5 eV, respectively, 
and still does not lead to a better quantitative agreement for the whole spectrum 
(see figure \ref{figure5}).

Introduction of the alternating-sign pseudopotential $\Delta U$ allows one to improve significantly 
the electron density distribution (see the dash-dotted blue curve in figure \ref{figure2}).
The difference between the {\it ab initio} calculated electron density and the one from the
jellium model calculation in the spatial region $8-12$ a.u. may contribute to the shift
of 6g and 7h ionization potentials (see figure \ref{figure5}).

\begin{figure}
\centering
\includegraphics[scale=0.5,clip]{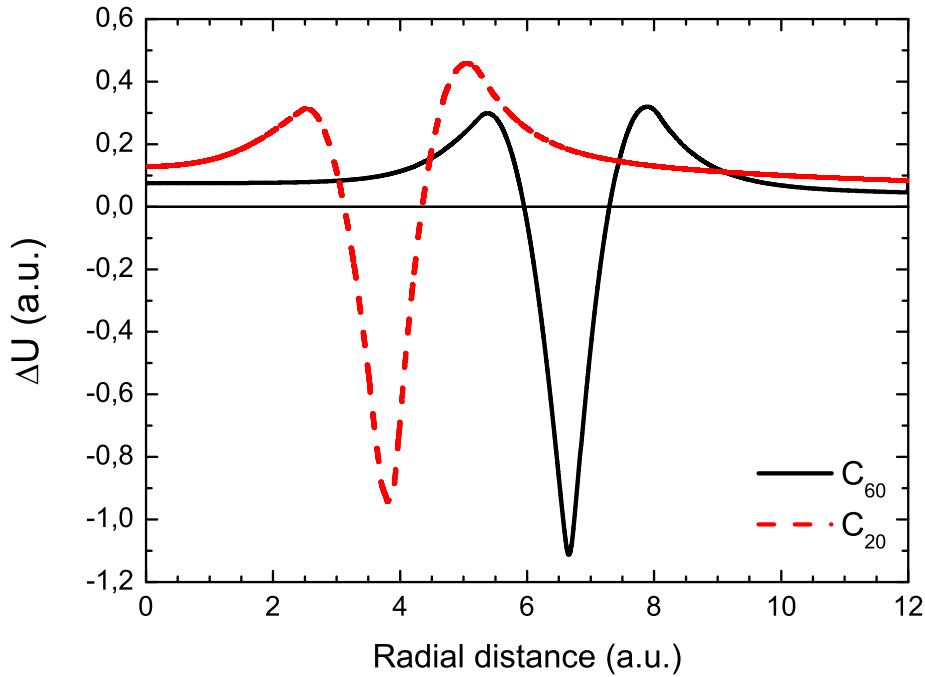}
\caption{
The additional pseudopotential $\Delta U$ in cases of C$_{60}$ (solid curve) 
and C$_{20}$ (dashed curve).} 
\label{figure9}
\end{figure}

Below we present and discuss the results for the C$_{20}$ molecule.
Following the formalism described above for C$_{60}$, the additional pseudopotential $\Delta U$ is
introduced as a difference between the total electrostatic potential of C$_{20}$ obtained from the
{\it ab initio} quantum-chemical calculation and the one obtained within the jellium model.
Figure \ref{figure9} represents the correction $\Delta U$ calculated for C$_{60}$ and C$_{20}$.
It is shown that $\Delta U$ has a similar alternating-sign shape for both molecules but it is
more asymmetric in the case of C$_{20}$.

\begin{figure}
\centering
\includegraphics[scale=0.35,clip]{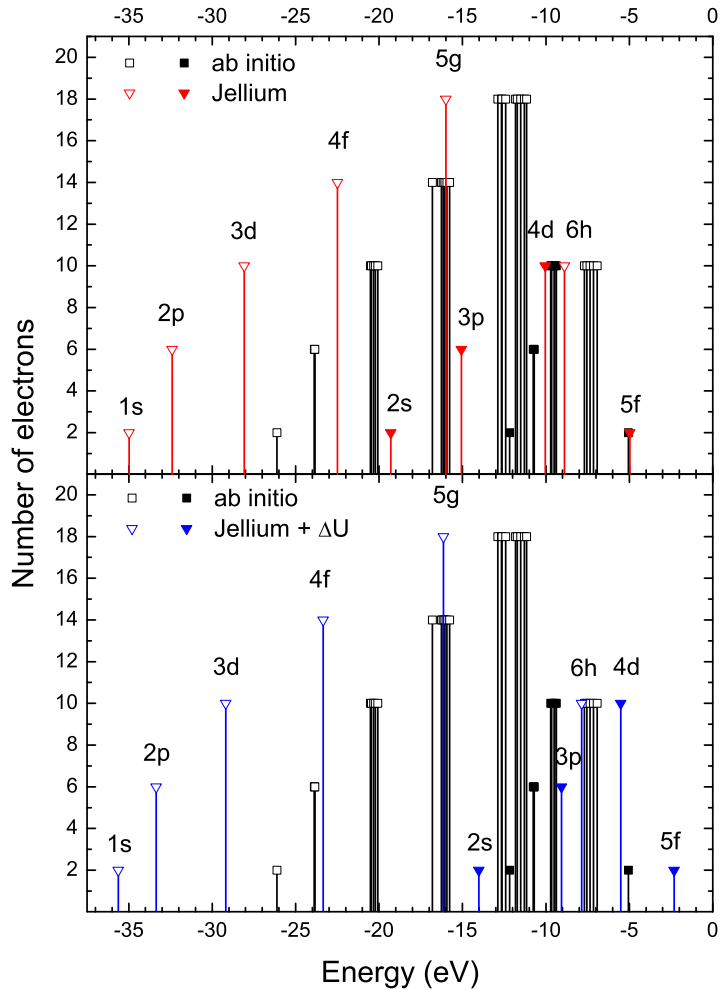}
\caption{
Single-electron energy levels of C$_{20}$ obtained from the {\it ab initio} calculation 
(empty and filled squares) and within the jellium model (empty and filled triangles): 
the standard one (upper panel) and the modified one (lower panel). 
Nodeless $\sigma$-orbitals and single-node $\pi$-orbitals are labeled by empty and filled symbols, 
respectively.}
\label{figure7}
\end{figure}

The single-electron energy spectra of C$_{20}$ are presented in figure \ref{figure7}.
The pseudopotential $\Delta U$ does not influence significantly on all nodeless orbitals 
while the single-node orbitals are shifted.
This shift leads to a better agreement of the {\it ab initio} and jellium calculations for
2s and 3p shells but gives a wrong value for the outer 4d and 5f ionization potentials.

\begin{figure}[h]
\centering
\includegraphics[scale=0.5,clip]{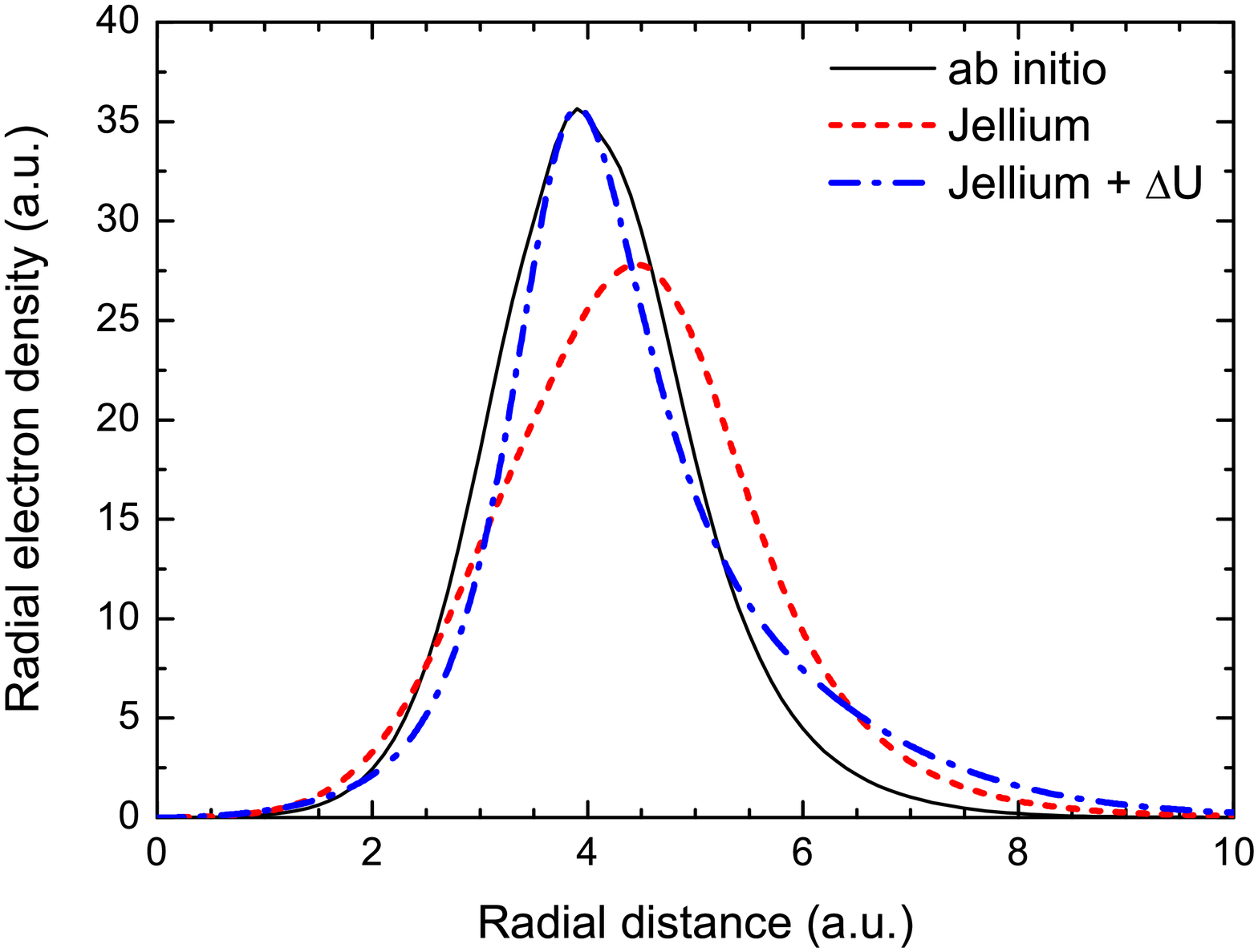}
\caption{
Radial electron density of C$_{20}$ obtained from the {\it ab initio} calculation (solid curve) 
and calculated within the jellium model, the standard one (dashed curve) and with the additional 
pseudopotential $\Delta U$ (dash-dotted curve).}
\label{figure8}
\end{figure}

The additional pseudopotential $\Delta U$ exerts a similar influence on the electron density 
distribution of C$_{20}$, as in case of C$_{60}$ (see figure \ref{figure8}).
In comparison with a standard jellium model (dashed red curve), the modified one improves the 
density distribution in the vicinity of the fullerene core (dash-dotted blue curve) but the 
electron density is spread partly to the spatial region $5-9$ a.u.

Having considered two different fullerenes within the spherical jellium model, one can conclude 
that the precise description of single-electron energy spectra of these systems by means of the 
jellium model is very difficult and elusive task, though such an approach produces mostly the 
right sequence of energy levels.
Additional pseudopotentials allow one to obtain the right value of the ionization potential only 
for several outer shells but do not alter the overall situation significantly.
At the same time, we suppose that by improving the ground-state density distribution with the 
introduced pseudopotential one can achieve higher accuracy while constructing the photoionization 
amplitudes.

\begin{figure}
\centering
\includegraphics[scale=0.35,clip]{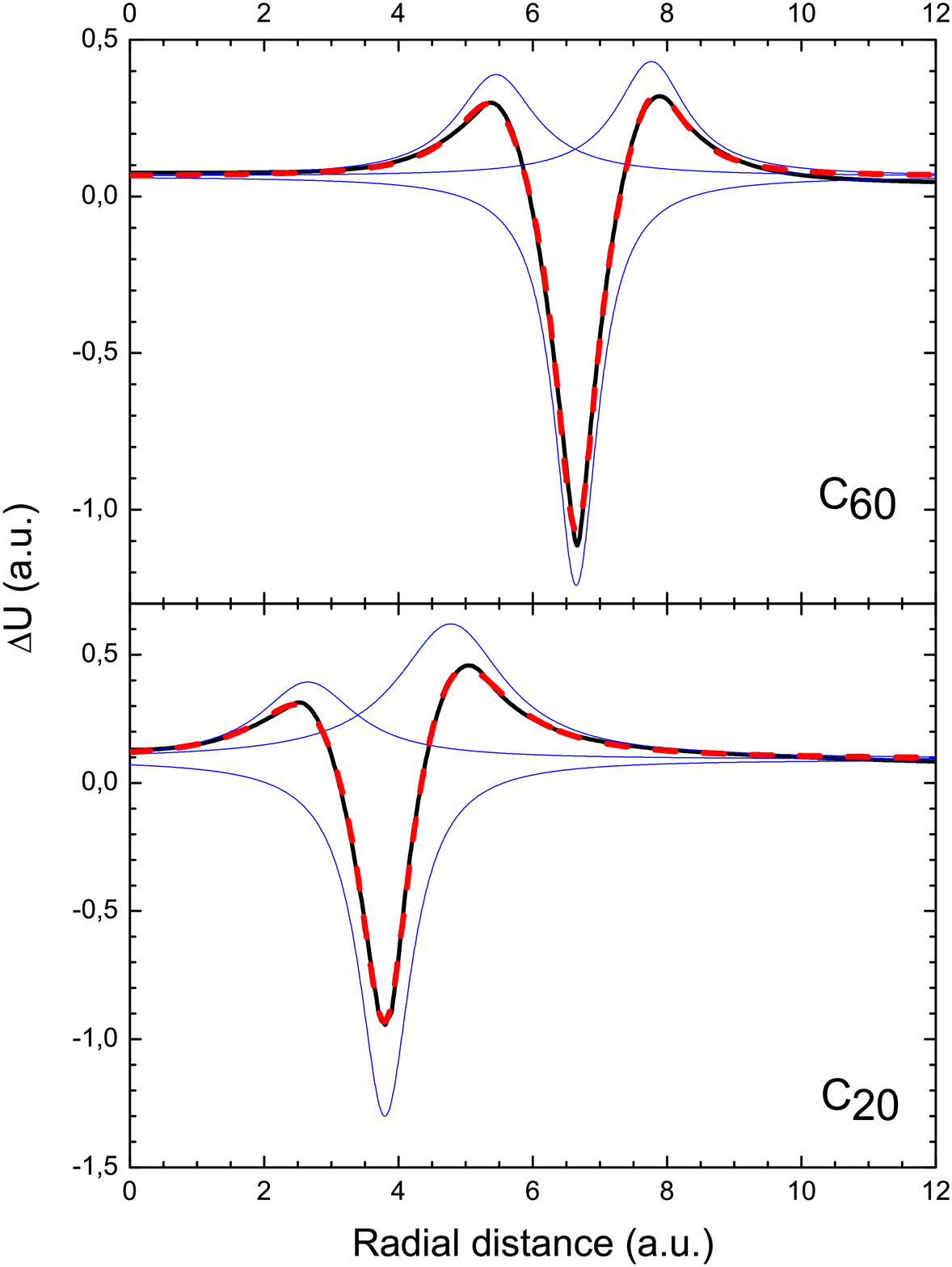}
\caption{
Pseudopotential $\Delta U$ for the C$_{60}$ (upper panel) and C$_{20}$ (lower panel) molecules. 
The initial curve is presented by the thick solid (black) line, dashed (red) line represents the 
fitting curve constructed as a sum of three primitive Lorentz functions (thin blue lines).}
\label{figure_fit}
\end{figure}

The obtained pseudopotentials for C$_{60}$ and C$_{20}$ can be well fitted by three Lorentz functions. 
The result of the fitting procedure is presented in figure \ref{figure_fit}.
Supposing $\Delta U(r) \equiv y(x)$, the resulting fitting function could be defined in the following form:
\begin{equation}
y(x) = y_0 + \sum_{i=1}^3 \frac{2A_i}{\pi} \frac{w_i}{4 (x - x_{c_i})^2 + w_i^2} \ ,
\end{equation}

\noindent where $y_0$ is the offset constant, $x_c$ is the position of the peak maximum,
$w$ is the full-width at half-maximum and $A$ is the normalization factor.
The obtained values of these parameters are presented in table~\ref{Table_1}.

\begin{table*}
\centering
\caption{Parameters of the Lorentz functions used for the fitting the pseudopotential $
\Delta U$ for C$_{60}$ and C$_{20}$.}
\begin{tabular}{|p{0.75cm}|p{1.1cm}|p{1.1cm}|p{1.1cm}|p{1.1cm}|p{1.1cm}|p{1.1cm}|p{1.1cm}|p{1.1cm}|p{1.1cm}|p{1.1cm}|}
\hline
         & $y_0$  & $x_{c_1}$ &  $w_1$  &  $A_1$  & $x_{c_2}$ &  $w_2$  &  $A_2$  & $x_{c_3}$ &  $w_3$  &  $A_3$  \\
\hline
C$_{60}$ & 0.064  &   5.453   &   1.425 &  0.727  &   6.647   &  0.785  & -1.610  &   7.763   &  1.264  &  0.727  \\
C$_{20}$ & 0.092  &   2.650   &   1.719 &  0.815  &   3.797   &  0.939  & -2.053  &   4.779   &  1.934  &  1.606  \\
\hline
\end{tabular}
\label{Table_1}
\end{table*}

\begin{figure}
\centering
\includegraphics[scale=0.35,clip]{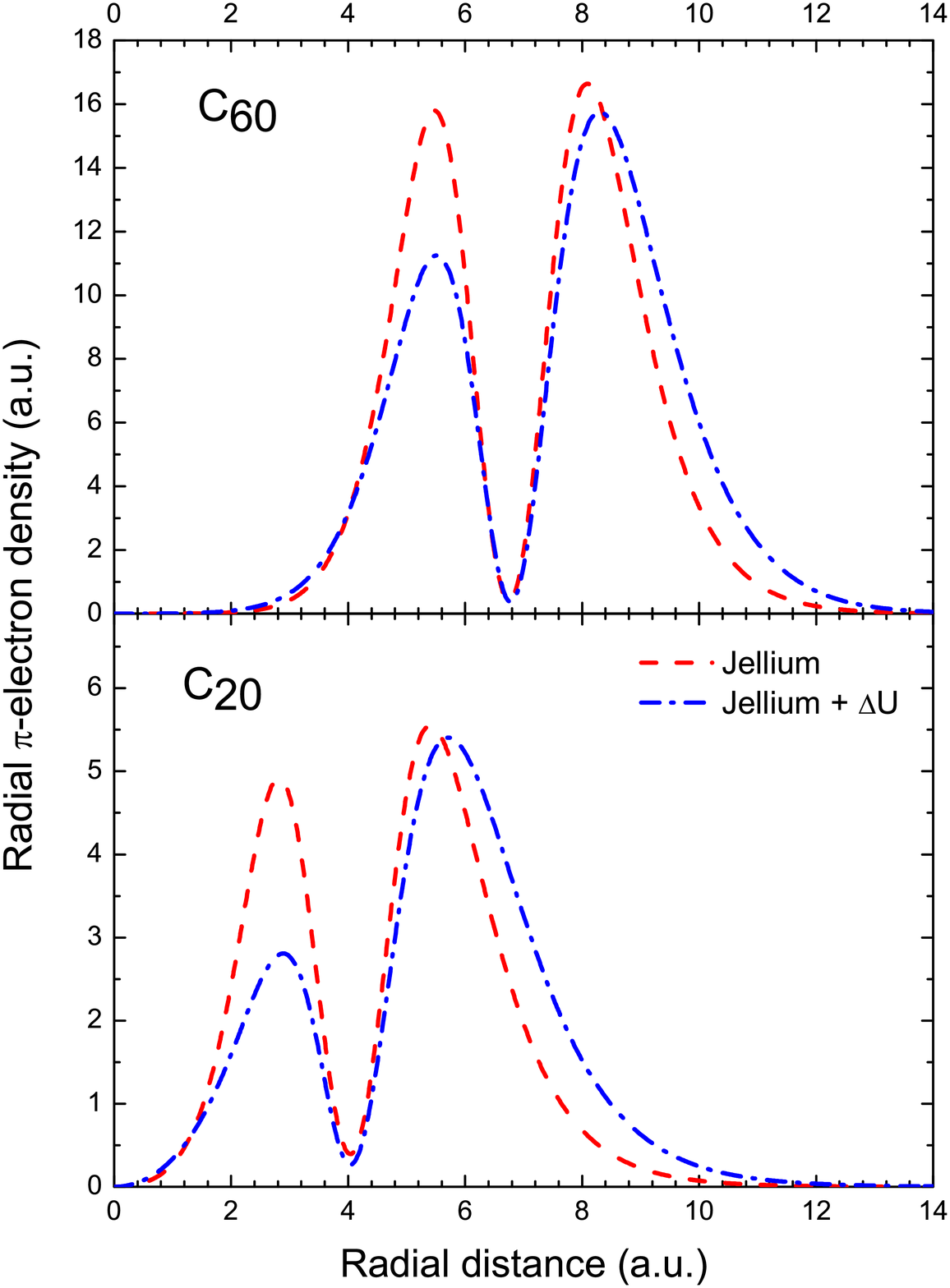}
\caption{
Radial density of $\pi$-electrons in C$_{60}$ (upper panel) and C$_{20}$ (lower panel) calculated 
within the standard jellium model (dashed red curve) and the modified jellium model with the presence 
of $\Delta U$ (dash-dotted blue curve).}
\label{figure10}
\end{figure}

As was shown above, the pseudopotential $\Delta U$ has a more asymmetric form in the case of C$_{20}$
than in the case of C$_{60}$; therefore, it should affect differently the $\pi$-electrons of these systems.
Figure \ref{figure10} represents the radial density of $\pi$-electrons in the C$_{60}$ and C$_{20}$
molecules obtained within the standard jellium model as well as the one augmented by $\Delta U$.
The minimum of the $\pi$-electron density distribution is located at 6.78 a.u. for C$_{60}$ and 4.03 a.u. 
for C$_{20}$.
These values are slightly shifted from the mean radius of the molecules, which equals 6.67 a.u. and
3.86 a.u., respectively.
It is shown that due to the hybridization-related correction $\Delta U$, $\pi$-electrons in both
systems are distributed non-uniformly in the inner and outer regions of the molecules.

\begin{figure}
\centering
\includegraphics[scale=0.5,clip]{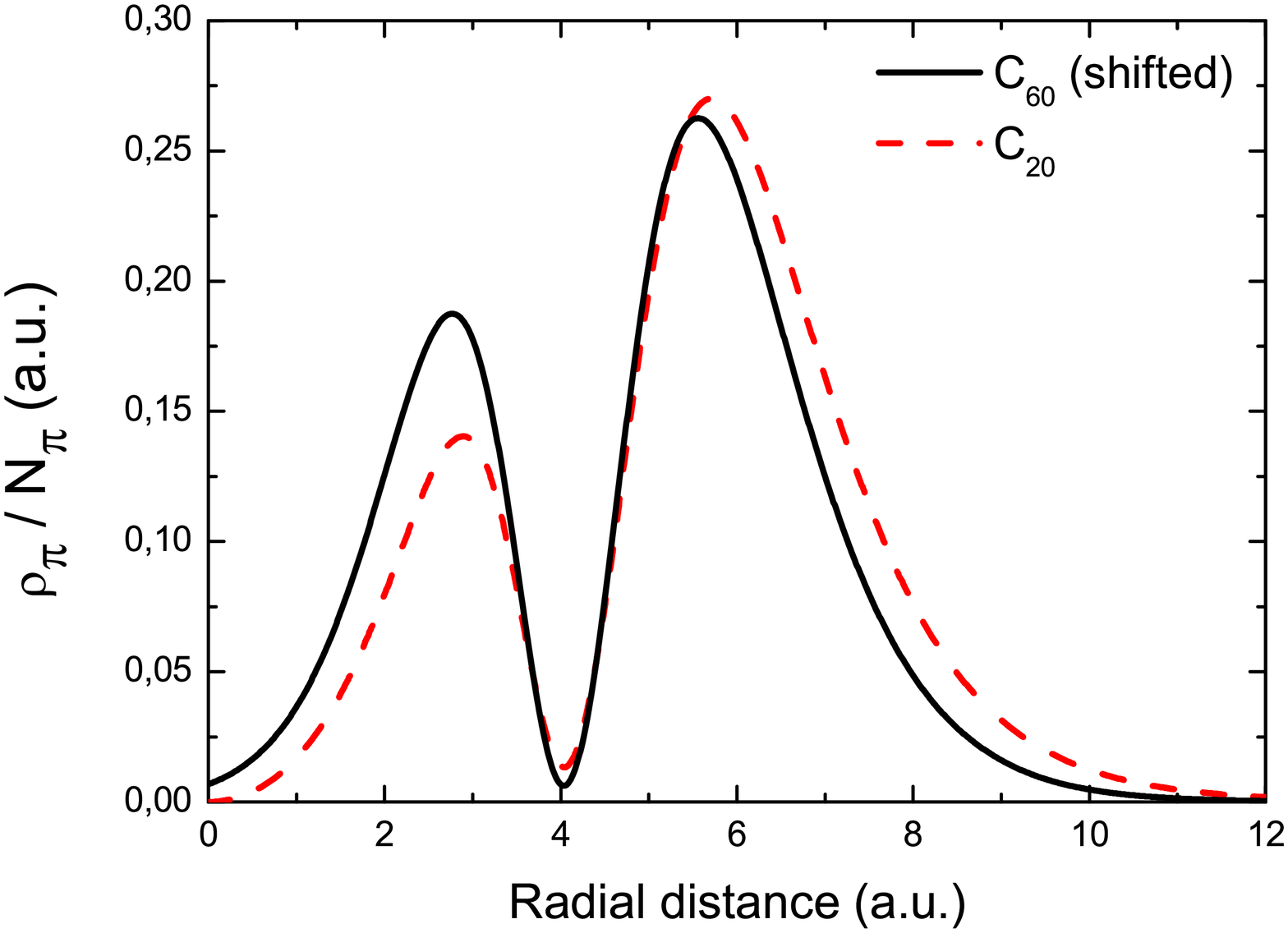}
\caption{
Radial $\pi$-electron densities, $\rho_{\pi}$, of the C$_{60}$ and C$_{20}$ fullerenes 
normalized by the number of the $\pi$-electrons, $N_{\pi}$, in each system. 
The density distribution of C$_{60}$ is shifted to match the minima of the two curves 
(see the text for more explanation).}
\label{figure11}
\end{figure}

To estimate a relative degree of spill-out of the $\pi$-electrons to the outer region of the 
fullerene molecules, we normalized the density distributions by dividing them by the number of 
the $\pi$-electrons in each system.
We also shifted the $\pi$-electron density of C$_{60}$ to the one of C$_{20}$ to match the minima
of the curves.
The result is presented in figure \ref{figure11}.
It is shown that the profile of the $\pi$-electron density in C$_{60}$ differs from the one in 
C$_{20}$.
Due to the smaller radius of the molecule and a bigger curvature of the fullerene surface,
$\pi$-electrons in C$_{20}$ are spilled out further than in the case of C$_{60}$.
On the basis of this comparison, we suppose that for larger fullerenes, such as C$_{240}$, 
$\pi$-electrons should be distributed more uniformly due to a smaller curvature of the surface 
of the molecule.

\section{Conclusion}

To conclude, we have introduced a new type of correction for description of the fullerenes 
C$_{60}$ and C$_{20}$ within the spherically symmetric jellium model.
The correction is represented as an additional pseudopotential which originates from the 
difference between the precise {\it ab initio} calculation and the one within the jellium model.
Due to the alternating-sign shape of the potential, it affects the $\sigma-$ and $\pi$-electrons 
of the system differently.
Therefore, this potential allows one to mimic partially the sp$^2$-hybridization, which occurs in 
formation of fullerenes, and, thus, to import the hybridization effects into the standard jellium model.
We have shown that the correction used improves significantly the electron density distribution as 
compared to the standard jellium model and the one with an additional square-well pseudopotential.
Like the other previously used corrections, it does not allow one to obtain a quantitative agreement 
with an {\it ab initio} calculation for the single-electron energy spectrum but reproduces the sequence 
of energy levels corresponding to the one following from the more precise quantum-chemical calculation.

As the next step of this work, the correction to the jellium model, introduced in this paper, will be 
utilized further for the calculation of the dynamic response of fullerenes in the processes of photon
and electron impact excitation.
We suppose that improving the ground-state density distribution with the introduced pseudopotential 
it is possible to get an accurate description of the excitation processes of fullerenes.
Particular attention will be paid on the study of collective electron excitations.
This work is currently in progress and the results will be presented elsewhere.
An implementation of the presented formalism for larger fullerene molecules, nanotubes etc. could be 
another topic of further investigations.

\section*{Acknowledgements}

A.V.V. is grateful to Deutscher Akademischer Austauschdienst (DAAD) for the financial support.
The authors acknowledge the Frankfurt Center for Scientific Computing for the opportunity of
carrying out complex {\it ab initio} calculations.
R.G.P. and V.K.I. thank Frankfurt Institute for Advanced Studies for hospitality and acknowledge 
support from Russian Federal Program 'Scientific and Educational Manpower for Innovative Russia' 
(grant No. 2012-1.5-12-000-1011-007).

\section*{References}


\end{document}